\documentclass[%
  aip,
  jcp,
  amsmath,amssymb,
  reprint,
  onecolumn
]{revtex4-1}

\usepackage{graphicx}
\usepackage{dcolumn}
\usepackage{bm}
\usepackage[utf8]{inputenc}
\usepackage[T1]{fontenc}
\usepackage{mathptmx}
\usepackage{etoolbox}
\usepackage{hyperref}
\usepackage{color}

\makeatletter
\def\@email#1#2{%
 \endgroup
 \patchcmd{\titleblock@produce}
  {\frontmatter@RRAPformat}
  {\frontmatter@RRAPformat{\produce@RRAP{*#1\href{mailto:#2}{#2}}}\frontmatter@RRAPformat}
  {}{}
}%
\makeatother

\newcommand{\ps}{p_{\rm s}}
\newcommand{\Qtot}{F_\Theta}
\newcommand{\Qself}{Q}
\newcommand{\ellzero}{\ell_0}

\newcommand{\mpc}[1]{{\color{black}{{#1}}}}

\begin{document}

\title{
Length-scale dependence of collective time reparameterization in swap dynamics
}

\author{Massimo Pica Ciamarra}
\email{massimo@ntu.edu.sg}
\affiliation{Division of Physics and Applied Physics, School of Physical and
Mathematical Sciences, Nanyang Technological University, Singapore 637371,
Singapore}
\affiliation{CNR--SPIN, Dipartimento di Scienze Fisiche, Universit\`a di Napoli
Federico II, I-80126 Napoli, Italy}

\date{\today}

\begin{abstract}
Swap Monte Carlo dramatically accelerates the equilibration of deeply
supercooled liquids while preserving the equilibrium Boltzmann distribution.
Recent work has argued that swap and other accelerated algorithms act by
reparameterizing time: different dynamics follow the same slow relaxation
process, but with different clocks. Here we test this idea in a continuously
polydisperse glass former. We confirm that mean collective overlaps satisfy an
approximate time reparameterization, in agreement with earlier results.
However, the associated clock depends systematically on the collective probe
length and on the relaxation stage. 
\mpc{Thus the observed collective collapse does not define a unique,
observable-independent material time. Over the numerically accessible
$\alpha$-relaxation regime, swap changes not only the rate of relaxation, but
also its length-scale dependence. The resulting collective
reparameterization is therefore only approximate and does not establish an
exact dynamical equivalence between swap and reference relaxation.}
\end{abstract}

\maketitle

\section{Introduction}
The origin of the rapid growth of relaxation times in glass-forming liquids
remains actively debated~\cite{berthier2011,dyre2006}. Upon cooling or
compression, structural relaxation slows by many orders of magnitude while
conventional static two-point correlations change only weakly. This decoupling
between structure and dynamics has motivated several theoretical frameworks.
Thermodynamic approaches, such as the Adam--Gibbs and random first-order
transition (RFOT) scenarios, tie the slowdown to a decreasing configurational
entropy and to growing amorphous order~\cite{adam1965,kirkpatrick1989,bouchaud2004,biroli2008,parisi2020}.
Kinetically constrained and facilitation-based descriptions instead emphasize
the dynamical organization of mobility~\cite{ritort2003,chandler2010,keys2011}.
Elastic and local-excitation approaches relate the slowdown to the evolution of
localized rearrangements, elastic properties, and their interactions~\cite{dyre2006,picaciamarra2026review}.
Structural approaches instead attribute the slowdown to the growth of subtle
local structural order that is largely invisible to conventional two-point
correlations yet correlates strongly with the emerging dynamic
heterogeneity~\cite{tanaka2010,tanaka2019}.
Distinguishing between these views is difficult because they often make similar
predictions for the temperature dependence of the relaxation time over the
numerically accessible range.

Swap Monte Carlo provides a particularly sharp test of these ideas. The method
augments ordinary translational moves with exchanges of particle diameters and,
for suitably chosen continuously polydisperse mixtures, accelerates
equilibration by many orders of magnitude while preserving the same equilibrium
measure~\cite{grigera2001,ninarello2017}. Since static observables and
thermodynamic quantities are unchanged whereas the dynamics is strongly
modified, swap dynamics directly probes which aspects of glassy relaxation are
controlled by equilibrium structure and which depend on the kinetic rules used
to explore configuration space.

One influential line of interpretation emphasizes this kinetic-rule sensitivity.
Wyart and Cates argued that the efficiency of swap dynamics limits the role
played by a growing static length in the ordinary glass transition and points
instead to local barriers or kinetic constraints as the dominant source of the
physical slowdown~\cite{wyart2017}. Related work generalized this idea by
continuously varying the kinetic rules of the same liquid: the apparent glass
transition and dynamical heterogeneity change strongly and smoothly with the
imposed kinetic constraints even though equilibrium structure is
unchanged~\cite{gavazzoni2024}. A complementary interpretation, proposed by
Berthier \textit{et al.}\ within the RFOT framework~\cite{berthier2019crumbling},
holds that swap can postpone the onset of amorphous metastability and thereby
reduce or bypass free-energy barriers in a regime where ordinary dynamics is
already activated. In this ``crumbling metastability'' picture, swap does not
simply multiply all microscopic rates by a constant; it opens relaxation
channels that destabilize metastable basins which are long lived under ordinary
dynamics. Both views predict that swap changes the character of relaxation,
not only the rate at which a fixed trajectory is traversed.

\mpc{
More recently, Ghimenti \textit{et al.}~\cite{ghimenti2026} proposed an
alternative interpretation: accelerated algorithms such as swap may act, in
the slow collective sector, primarily through a reparameterization of time.
In its strongest form, this picture implies that different dynamics traverse
the same slow relaxation trajectory with different clocks. A genuine
single-clock description then requires the same time transformation to apply
to all collective observables.
}

\mpc{
Here we test this requirement explicitly across spatial probe lengths. We first
confirm that mean collective overlaps exhibit the approximate parametric
collapse reported in Ref.~\citenum{ghimenti2026}. We then extract the material
clock independently from different probe lengths and find that it depends
systematically on both probe length and relaxation stage. A complementary
distribution-level analysis shows that the underlying nearest-distance
distribution also remains swap dependent at matched collective relaxation
stages. 
Thus, for the model and relaxation regime studied here, the collective
collapse is only a projected approximation and does not define an exact,
observable-independent material clock. Swap therefore changes more than
the pace at which an otherwise fixed slow relaxation process is traversed,
leaving unresolved the tension between landscape-based and kinetic
descriptions of glassy dynamics.}

\mpc{
\section{Theoretical Background: From Aging Dynamics to Accelerated Sampling}
\label{sec:theoretical_background}
Time-reparameterization invariance (TRI) was originally introduced in the field 
theory of out-of-equilibrium aging systems~\cite{Chamon2002,Kennett2001}, where 
Time-Translation Invariance (TTI) is broken and two-time correlation functions 
$C(t_a, t_b)$ depend explicitly on both times rather than solely on their difference. 
In that framework, continuous time reparameterization, $t \to h(t)$, describes the 
slow, structural $\alpha$-relaxation sector once a pronounced separation between 
microscopic $\beta$ and structural timescales exists ($\tau_\alpha / \tau_\beta \gg 1$). 
Crucially, TRI is predicted to hold across the entire $\alpha$ and late-$\alpha$ 
relaxation regime, after fast, short-time vibrational 
motion decays and ceases to influence the dynamics.
In the field-theoretic formulation, TRI
is a symmetry of the slow action, predicting that a single, 
universal material clock $h(t)$ governs all slow two-time observables—including 
both self (single-particle) and collective density correlations. }

\mpc{
Strictly speaking, non-trivial time reparameterizations do not exist in standard 
equilibrium states: enforcing TTI requires $C(t_a, t_b) = C(t_a - t_b) = C(h(t_a) - h(t_b))$, 
which fixes the reparameterization gauge down to linear time shifts, 
$h(t_a) - h(t_b) = t_a - t_b$, implying $h(t) = t + \text{const}$.}

\mpc{
The conceptual framework introduced by Ghimenti \textit{et al.} in Ref.~\citenum{ghimenti2026} 
corresponds to an extension of this continuous clock transformation to equilibrium 
dynamics under accelerated algorithms, effectively mapping the algorithmic parameter 
space onto the aging time domain. 
Specifically, for the swap-accelerate dynamics, the swap probability $p_s$ plays a role 
analogous to the waiting time $t_w$ in aging: a smaller swap probability $p_s \to 0$ 
corresponds to an ``older,'' slower physical system, whereas a higher swap probability $p_s > 0$ acts as a dynamical acceleration. }

\mpc{
In direct analogy with the aging formulation, this reparameterization is predicted 
to hold across the structural $\alpha$-relaxation sector after the fast microscopic 
$\beta$ regime has decayed. The central hypothesis of this extension is that varying 
$p_s$ merely rescales the material clock via a single scalar transformation,
\begin{equation}
t \to h_{p_s}(t),
\end{equation}
leaving the underlying physical relaxation trajectory across all spatial modes intact. 
Under a genuine single-clock reparameterization, all slow observables within a given 
sector must transform under the same change of time. }

\mpc{
In finite-dimensional particle liquids under swap Monte Carlo, full field-theoretic 
time-reparameterization invariance fails in both the fluctuation and self sectors. 
Because soft reparameterization modes generate dynamic heterogeneity, exact TRI 
demands a universal parametric collapse for four-point susceptibilities $\chi_4$, 
which swap is well known to violate by strongly suppressing spatial mobility correlations. 
Furthermore, self observables strongly reflect local, algorithm-specific diameter 
exchanges that bypass collective cage restructuring, causing single-clock collapse to 
fail severely for single-particle motion (see Appendix~\ref{app:beyond_mean}). 
}

\mpc{
The collective mean-overlap sector therefore represents the most favorable, minimal 
regime where an effective single-clock reparameterization could plausibly hold, 
making it the primary testing ground for the framework.
If TRI holds, then for each choice of swap probability $p_s$, there 
exists a single monotonic material clock $h_{p_s}(t)$ such that the collective overlap 
$F_\Theta(\hat{a}, t)$ probed at any length scale $\hat{a}$ satisfies
\begin{equation}
F_\Theta^{(p_s)}(\hat{a}, t) \simeq F_\Theta^{({\rm ref})}(\hat{a}, h_{p_s}(t))
\label{eq:single_clock_intro}
\end{equation}
with the exact same $h_{p_s}(t)$ across all probe lengths $\hat{a}$. 
}

\mpc{
In this work,  we systematically test its mapping across the operational $\alpha$-relaxation window in the collective mean-overlap sector. By explicitly extracting $h_{p_s}^{(\hat{a})}(t)$ for different probe length scales $\hat{a}$, we evaluate whether the swap-induced speedup can indeed be understood as a unique, scalar material time reparameterization, or whether swap Monte Carlo fundamentally alters the relaxation mechanisms across spatial modes.}

\section{Model and methods}
We study the three-dimensional non-additive polydisperse glass former of
Ref.~\citenum{Lerner2019}. Particle diameters follow $P(\sigma)\propto\sigma^{-3}$
over a range $[\sigma_{\min},\sigma_{\max}]$ with
$\sigma_{\max}/\sigma_{\min}=2.2$, and particles interact through the purely
repulsive potential
$V(r)=\epsilon\left(\sigma_{ij}/r\right)^{10}+s(\sigma_{ij}/r)$, where $s(x)$ is
a smoothing function ensuring continuity up to three derivatives at the cutoff
$x=1.4$, and $\sigma_{ij}$ is a non-additive interaction
diameter~\cite{Lerner2019}. We take $\sigma_{\min}=1$ and $\epsilon$ as the
units of length and energy, and simulate $N=2000$ particles at number density
$\rho=0.58$ in a periodic cubic box, preparing equilibrium configurations with
the swap algorithm. 
The main tests are performed at $T=0.50$, corresponding to the critical
temperature for this model~\cite{cia24}. This is the lowest temperature at which
the ordinary (weak-swap) dynamics relaxes fully over the accessible simulation
window. 
We include additional high-overlap checks at $T=0.45$, at which it is not currently possible to follow the full dynamical relaxation curves.
\mpc{At $T=0.50$, averages are obtained from 100 independent trajectories for each
dynamics, initialized from independently equilibrated configurations.}

The dynamics is sampled by Monte Carlo. In one Monte Carlo sweep we attempt $N$
elementary displacement moves and $N\ps$ diameter-swap moves between randomly
selected particles. All moves obey detailed balance with
respect to the same equilibrium Boltzmann distribution, so changing $\ps$
changes the kinetic rules but not the target equilibrium measure.

Throughout the analysis, all particle displacements are computed after
subtracting the center-of-mass displacement between the two configurations
being compared. We monitor collective density relaxation through the normalized
overlap
\begin{equation}
\Qtot(a,t)=\frac{F(a,t)-q_r}{1-q_r},
\qquad
F(a,t)=\frac{1}{N}\sum_{i,j}
\theta\!\left(a-|\mathbf r_i(t)-\mathbf r_j(0)|\right),
\label{eq:collective_overlap}
\end{equation}
where $q_r=(4\pi/3)\rho a^3$ is the random-overlap contribution in three
dimensions. This observable compares coarse-grained density fields at two times
and does not require particle identity to be preserved. 
For the overlap probe length, we report distances in units of the typical interparticle spacing $\ellzero=\rho^{-1/3}$ and define $\hat a=a/\ellzero$.

We quantify the fluctuations of the collective overlap through the four-point
susceptibility
\begin{equation}
\chi_\Theta(a,t)=N\left(\langle F_\Theta(a,t)^2\rangle
-\langle F_\Theta(a,t)\rangle^2\right),
\label{eq:chi4}
\end{equation}
computed from the variance of $F_\Theta$ over independent realizations. 

\mpc{Statistical uncertainties in the clock and derived quantities are estimated
by bootstrap resampling at the trajectory level. Each bootstrap realization is
constructed by resampling the complete set of independent trajectories with
replacement, so that correlations among times and probe lengths within a given
trajectory are preserved. For each resample, the mean overlap curves are
recomputed and the full subsequent analysis, including interpolation, inversion
of the relaxation curves, and construction of the material clocks, is repeated.
The resulting distribution of each derived quantity is used to determine the
bootstrap confidence bands shown in the figures; the number of bootstrap
resamples and the confidence level displayed are specified in the corresponding
figure captions.}

To probe the microscopic route to relaxation, we also consider the self-overlap
\begin{equation}
\Qself(a,t)=\frac{1}{N}\sum_{i=1}^{N}
\exp\!\left[-\frac{|\mathbf r_i(t)-\mathbf r_i(0)|^2}{2a^2}\right],
\label{eq:self_overlap}
\end{equation}
and the self part of the van Hove distribution,
\begin{equation}
G_s(\Delta r;t)=\frac{1}{3N}\sum_{i=1}^{N}\sum_{\alpha=x,y,z}
\left\langle
\delta\!\left(\Delta r-\left|r_{i\alpha}(t)-r_{i\alpha}(0)\right|\right)
\right\rangle .
\label{eq:vanhove}
\end{equation}

\section{Results}
\subsection{Mean collective overlaps approximately reparameterize}
\label{subsec:collective}

\begin{figure}[!t]
  \includegraphics[width=1.0\textwidth]{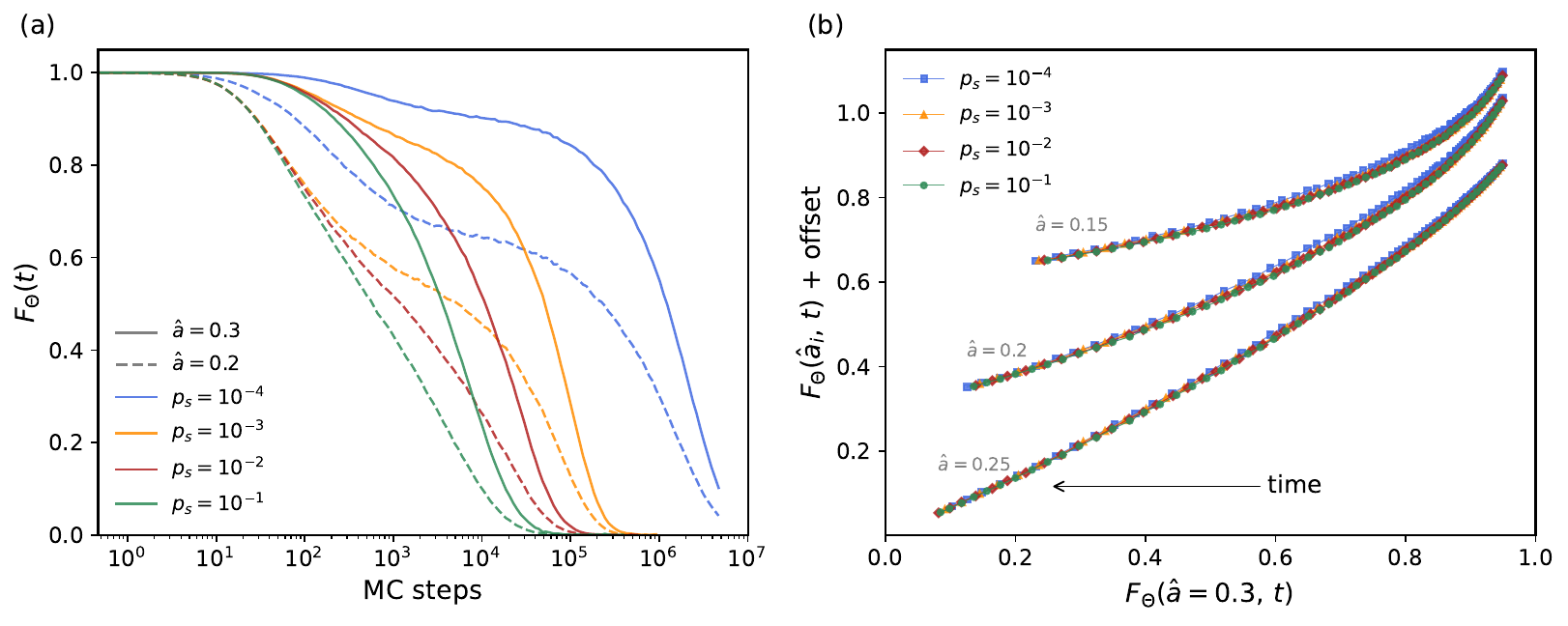}
  \caption{(a) Collective overlap $\Qtot(\hat a,t)$ as a function of Monte
  Carlo sweeps for probe lengths $\hat a=0.3$ (solid) and $\hat a=0.2$
  (dashed), and several swap probabilities $\ps$ (colors). Increasing $\ps$
  accelerates the decay. (b) Parametric plots of $\Qtot(\hat a_i,t)$ versus
  the coarse reference $\Qtot(\hat a=0.3,t)$, for probe lengths
  $\hat a_i=0.15,\,0.20,\,0.25$ and several values of $\ps$. Curves are
  shifted vertically for clarity. The approximate collapse reproduces the
  collective-overlap reparameterization reported in Ref.~\citenum{ghimenti2026}.}
  \label{fig:Ft}
\end{figure}

Figure~\ref{fig:Ft}(a) shows the collective overlap $\Qtot(\hat a,t)$ for two
probe lengths and several values of the swap probability, at $T=0.50$.
Increasing $\ps$ strongly accelerates relaxation, shifting the decay curve over
roughly two decades in Monte Carlo time.

The corresponding parametric plots are shown in Fig.~\ref{fig:Ft}(b). Plotting
$\Qtot(\hat a_i,t)$ against the reference $\Qtot(0.3,t)$ eliminates time; the
curves obtained at different $\ps$ collapse to a good approximation, with the
largest deviations at the smallest swap probability and the smallest probing
length. This reproduces the empirical observation of Ref.~\citenum{ghimenti2026}:
in the collective-overlap channel, the mean relaxation can be approximately
described by a reparameterization of time. We now ask a stricter question: does
this collapse define a unique material clock?

\subsection{Single-clock reparameterization requires a unique material time}
\label{subsec:singleclock_def}

A genuine time reparameterization requires more than the collapse of one
correlation function. For each dynamics, labelled here by the swap probability
$\ps$, there should exist a single monotonic clock $h_{\ps}(t)$ such that
\begin{equation}
\Qtot^{(\ps)}(\hat a,t)
\simeq
\Qtot^{({\rm ref})}(\hat a,h_{\ps}(t))
\label{eq:single_clock}
\end{equation}
for every probe length $\hat a$. Here the reference dynamics is chosen as the
smallest swap probability considered, $\ps^{\rm ref}=10^{-4}$, which is the
smallest swap rate for which the reference dynamics fully relaxes within the
time window accessible at $T=0.50$. Equation~\eqref{eq:single_clock}
therefore states that changing the swap probability changes the rate at which
the relaxation is traversed, while a unique material time describes all probe
lengths.

\mpc{
We test this condition by extracting the clock independently at each probe
length. For a given swap dynamics $p_s$ and physical time $t$, we define
$t_{\rm ref}^{(\hat a)}(t)$ as the reference time satisfying
\begin{equation}
\Qtot^{({\rm ref})}\!\left(\hat a,t_{\rm ref}^{(\hat a)}(t)\right)
=
\Qtot^{(\ps)}(\hat a,t).
\label{eq:clock_match}
\end{equation}
The corresponding material clock is
\begin{equation}
h_{\ps}^{(\hat a)}(t)
\equiv
t_{\rm ref}^{(\hat a)}(t)
=
\left[\Qtot^{({\rm ref})}(\hat a,\cdot)\right]^{-1}
\left[\Qtot^{(\ps)}(\hat a,t)\right].
\label{eq:clock_extract}
\end{equation}
Thus, for a fixed $p_s$ and physical time $t$, all probe-dependent clocks are
constructed from the same point along the swap trajectory; only the probe
length used to determine the corresponding reference time is changed.
If a single material clock exists, $h_{\ps}^{(\hat a)}(t)$ must be independent
of $\hat a$. Operationally, for the inversion, the reference relaxation curve at each probe length is replaced by a monotonic envelope to remove small statistical non-monotonicities.
}

\mpc{
We quantify the non-uniqueness of the material clock using two complementary
measures. The first directly compares the clock inferred at probe length
$\hat a$ with that obtained from the reference probe $\hat a_0=0.30$. The
second quantifies the consequence of this clock difference for the relaxation
function itself.
}

The direct clock difference is
\begin{equation}
\Delta \log_{10} h_{\ps}^{(\hat a)}
=
\log_{10} h_{\ps}^{(\hat a)}
-
\log_{10} h_{\ps}^{(\hat a_0)}.
\label{eq:clock_difference}
\end{equation}
This quantity measures the disagreement between the two clocks in
logarithmic-time units. A single-clock reparameterization requires
$\Delta \log_{10} h_{\ps}^{(\hat a)}=0$ for every $\hat a$.

\mpc{
To motivate the second measure, consider the collective overlap at probe
length $\hat a$. At a given physical time $t$, the clock
$h_{\ps}^{(\hat a)}(t)$ identifies the corresponding point along the
reference relaxation. Using instead the material time
$h_{\ps}^{(\hat a_0)}(t)$ inferred from the reference probe gives a second
prediction for the same overlap. We therefore define
\begin{equation}
\delta F_{\rm equiv}^{(\hat a)}(t)
=
\left|
\Qtot^{({\rm ref})}
\!\left(\hat a,h_{\ps}^{(\hat a)}(t)\right)
-
\Qtot^{({\rm ref})}
\!\left(\hat a,h_{\ps}^{(\hat a_0)}(t)\right)
\right|.
\label{eq:deltaF_equiv}
\end{equation}
This quantity directly measures the effect of the clock mismatch on the
relaxation function. If the two material times are identical,
$\Delta\log_{10}h_{\ps}^{(\hat a)}=0$, the two predictions coincide and
$\delta F_{\rm equiv}^{(\hat a)}=0$.

For small clock differences, this relation can be expressed locally as
\begin{equation}
\delta F_{\rm equiv}^{(\hat a)}
\simeq
\left|
\frac{d\Qtot^{({\rm ref})}(\hat a,h)}
     {d\log_{10}h}
\right|
\left|
\Delta\log_{10}h_{\ps}^{(\hat a)}
\right|.
\label{eq:deltaF_equiv_linear}
\end{equation}
The local slope of the reference relaxation curve therefore converts a
difference in material time into the corresponding difference in overlap.
A given clock mismatch can consequently have different dynamical significance
at different stages of the relaxation.}

\mpc{
The two measures thus characterize different aspects of the same violation of
a unique material time. $\Delta\log_{10}h_{\ps}^{(\hat a)}$ measures the
difference between the clocks themselves, whereas
$\delta F_{\rm equiv}^{(\hat a)}$ measures the resulting difference in the
relaxation function. The latter is particularly useful when comparing
different stages of the relaxation, since the same difference in logarithmic
material time can correspond to very different changes of the correlation
function. 
}
\mpc{
We perform these comparisons over the common relaxation window
\begin{equation}
0.2 \leq \Qtot^{(\ps)}(\hat a_0,t) \leq 0.8,
\qquad \hat a_0=0.30.
\label{eq:clock_window}
\end{equation}
This choice is physically motivated by the regime in which time
reparameterization is expected to apply. The time-reparameterization scenario
concerns the slow structural dynamics and is therefore expected, if applicable,
to emerge in the $\alpha$-relaxation regime, after microscopic and
$\beta$-relaxation processes have become subdominant and before the
correlation function becomes too close to its equilibrium limit. The interval
in Eq.~\eqref{eq:clock_window} selects this regime while retaining a common
relaxation window for all swap probabilities and probe lengths. Tracking the
two clock mismatches across this interval then tests both whether a unique
material time describes the $\alpha$ relaxation and whether such a
description becomes progressively more accurate as the relaxation proceeds.
}

\subsection{Probe-dependent clocks reveal non-uniqueness}
\label{subsec:clock_ratios}

\begin{figure}[!t]
  \includegraphics[width=1.0\textwidth]{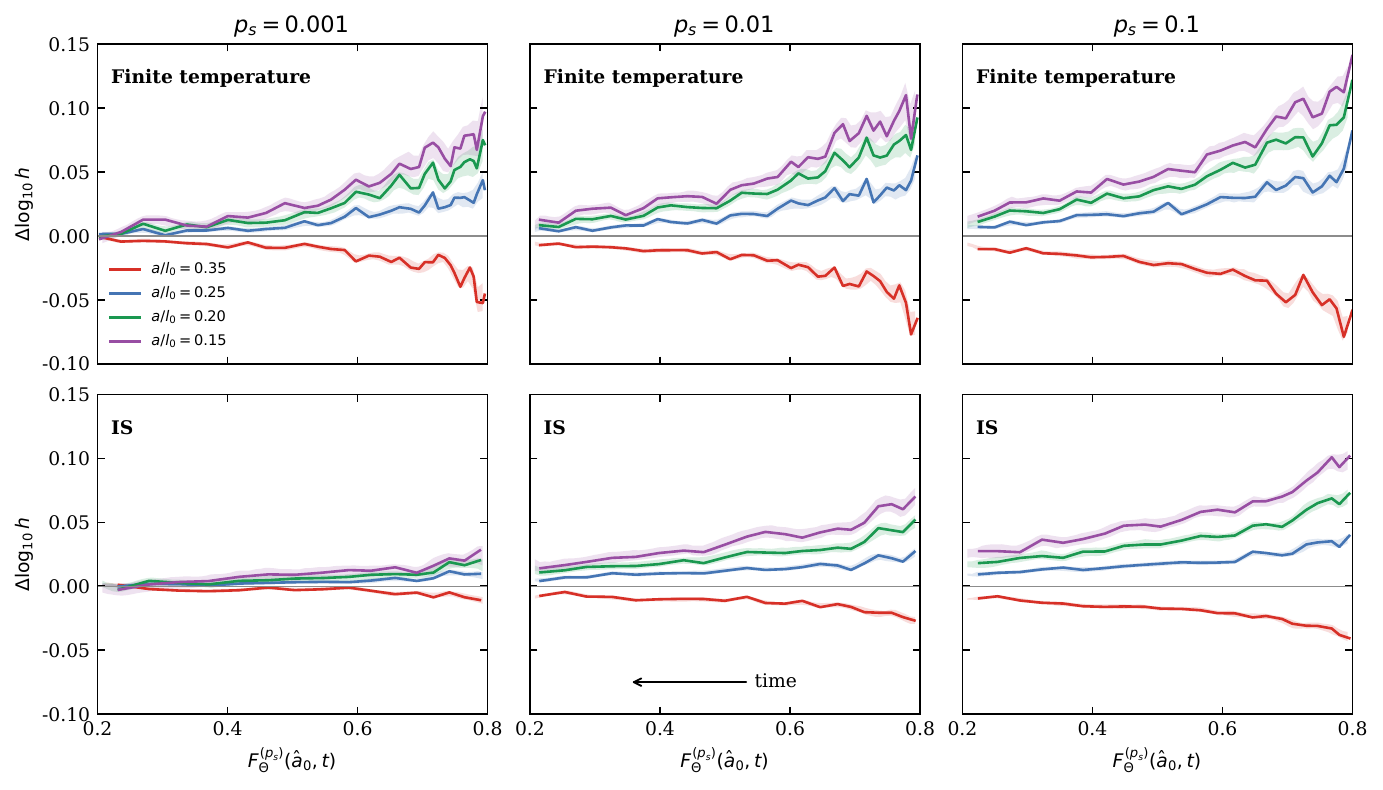}
  \caption{Signed clock differences at $T=0.50$ for
  $\ps=10^{-3}$, $10^{-2}$, and $10^{-1}$, shown for thermal trajectories
  (top row) and their inherent-structure projections (bottom row).
  For each swap probability, the clocks $h_{\ps}^{(\hat a)}(t)$ are extracted
  independently at different probe lengths and compared with the clock obtained
  from the reference probe $\hat a_0=0.30$.
  The plotted quantity is
  $\Delta\log_{10}h=
  \log_{10}h_{\ps}^{(\hat a)}-
  \log_{10}h_{\ps}^{(\hat a_0)}$,
  shown as a function of the relaxation stage
  $\Qtot(\hat a_0,t)$.
  Shaded bands indicate the central 68\% bootstrap confidence interval obtained
  from 300 trajectory-level resamples.
  A strict single-clock reparameterization would give
  $\Delta\log_{10}h=0$ for all probe lengths. Instead, the deviations are
  systematic, ordered by probe length, and vary across the relaxation.}
  \label{fig:signed_clock_T050}
\end{figure}
Figure~\ref{fig:signed_clock_T050} reports the signed clock differences at
$T=0.50$ for three swap probabilities, $\ps=10^{-3}$, $10^{-2}$, and
$10^{-1}$. The top row shows the results obtained from the thermal trajectories.
For each $\ps$, the clocks extracted at different probe lengths are compared
with the clock obtained from the reference probe $\hat a_0=0.30$, and the
difference is shown as a function of the relaxation stage
$\Qtot(\hat a_0,t)$.

The first observation is that the magnitude of the discrepancy depends
systematically on the swap probability. It is smallest for $\ps=10^{-3}$ and
becomes progressively larger as $\ps$ increases. This is the expected trend:
for weak swap dynamics the evolution remains close to the reference dynamics,
whereas increasing $\ps$ produces a larger modification of the relaxation and
therefore a more stringent test of whether that modification can still be
described by a single change of clock.

More importantly, the discrepancy is not constant during the relaxation. For
a given $\ps$ and probe length, $\Delta\log_{10}h$ varies systematically as
the system advances through the $\alpha$-relaxation regime. A strict
time-reparameterization description would instead require all probe-dependent
clocks to coincide throughout this regime. The finite and relaxation-stage
dependent differences observed in the top row therefore demonstrate that the
thermal dynamics is not related to the reference dynamics by a unique material
clock over the probed $\alpha$-relaxation window, despite the visually good
collapse of the mean collective overlaps.

The probe-length dependence is itself systematic. Relative to the reference
probe $\hat a_0=0.30$, smaller probe lengths give clock differences of one
sign, while the larger probe $\hat a=0.35$ gives the opposite sign. The clocks
are therefore not scattered randomly around the reference one, but are ordered
with probe length. This shows that the departure from a unique clock reflects
a genuine length-scale dependence of the swap-induced acceleration. In
particular, the acceleration inferred from finer probes differs systematically
from that inferred from coarser probes.
Appendix~\ref{app:globalclock} shows that the same conclusion is reached when
a single clock is optimized to describe all probe lengths simultaneously.

\mpc{
One possible concern is that this violation could originate from an imperfect
separation between the slow structural relaxation and faster vibrational or
intra-basin motion. To test this possibility, we repeated the analysis after
projecting the trajectories onto their inherent structures, a standard way of
suppressing the contribution of vibrational motion. The corresponding clock
differences are shown in the bottom row of
Fig.~\ref{fig:signed_clock_T050}. The same qualitative features remain:
the clocks depend systematically on probe length, the discrepancy increases
with $\ps$, and the differences remain finite across the slow relaxation.
The violation of a unique material clock is therefore not removed by filtering
out vibrational motion.}

The magnitude of the mismatch remains modest compared with the overall
acceleration produced by swap, but it is not negligible for a supposedly
unique material time. For example,
$\Delta\log_{10}h=0.1$ corresponds to a factor
$10^{0.1}\simeq1.26$ in material time. The bootstrap confidence intervals show
that the probe-dependent differences are statistically resolved over a
substantial part of the relaxation window. 
\mpc{In the following, we quantify not
only the difference between the extracted clocks themselves, but also the
corresponding difference they imply for the relaxation functions.}

\subsection{Clock mismatch in relaxation-function units}
\label{subsec:normalized_clock}
\mpc{
The probe-dependent clocks in Fig.~\ref{fig:signed_clock_T050} provide the
direct test of time reparameterization: a unique material clock would require
the same $h_{\ps}(t)$ for all probe lengths. The observed
$\Delta\log_{10}h\neq0$ therefore establishes the violation of a strict
single-clock description.
The magnitude of a clock difference in logarithmic-time units, however, does
not by indicate how strongly that violation affects the relaxation
functions. A given horizontal displacement in material time can correspond to
a large or small vertical change of the overlap depending on the local slope
of the relaxation curve. We therefore also express the same clock mismatch in
observable space through $\delta F_{\rm equiv}^{(\hat a)}$,
Eq.~\eqref{eq:deltaF_equiv}.
}

\begin{figure}[!t]
  \includegraphics[width=1.0\textwidth]{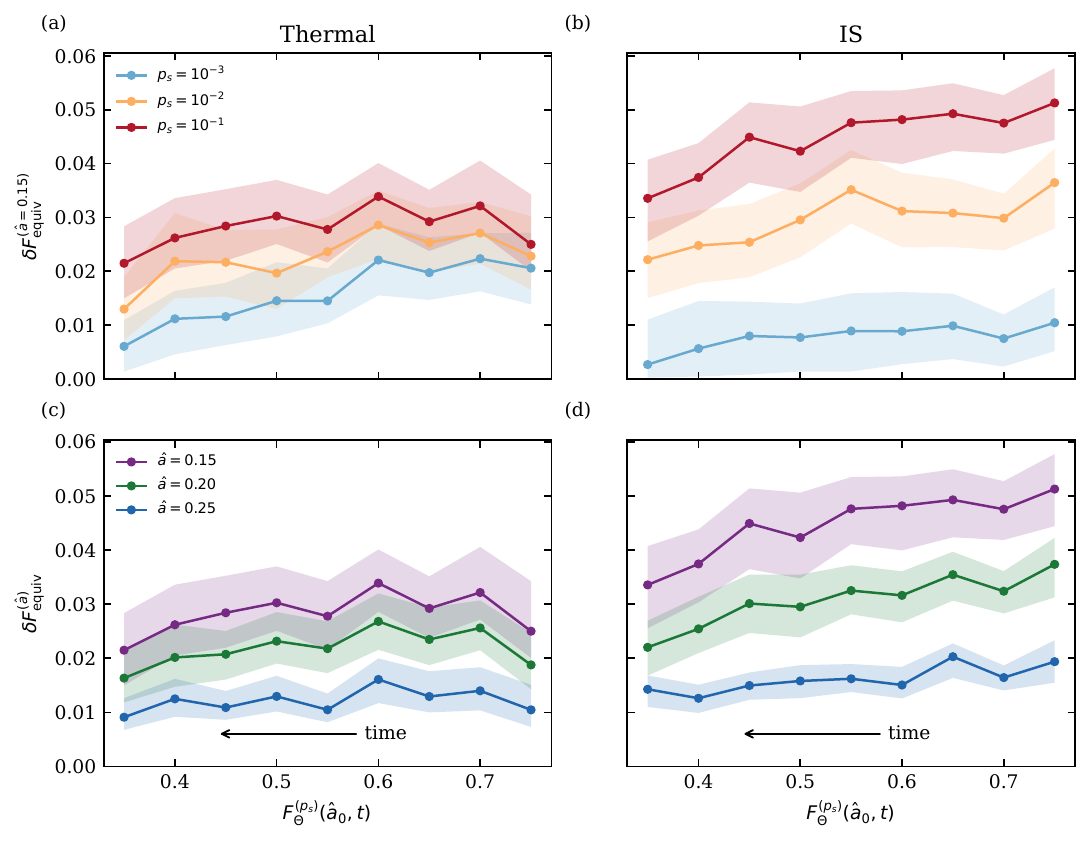}
  \caption{Dynamical significance of the probe-dependent clock mismatch at
  $T=0.50$, quantified by
  $\delta F_{\rm equiv}^{(\hat a)}$, Eq.~\eqref{eq:deltaF_equiv}.
  The relaxation stage is parametrized by
  $F=\Qtot^{(\ps)}(\hat a_0,t)$, with $\hat a_0=0.30$.
  Top row: dependence on swap probability at fixed probe length
  $\hat a=0.15$, for the thermal trajectories (a) and their
  inherent-structure projections (b).
  Bottom row: dependence on probe length at fixed strong swap,
  $\ps=10^{-1}$, for the thermal trajectories (c) and inherent structures
  (d). The reference-probe value is identically zero and is not shown.
  Shaded bands indicate the central 95\% bootstrap confidence interval obtained
from 200 trajectory-level resamples.
  The mismatch increases with swap probability and depends systematically on
  probe length. It remains finite throughout the resolved
  $\alpha$-relaxation window, both before and after projection onto inherent
  structures.}
  \label{fig:deltaF_equiv}
\end{figure}

\mpc{
We first examine how the dynamical effect of the clock mismatch depends on the
strength of the swap dynamics. 
Figure~\ref{fig:deltaF_equiv}(a) shows
$\delta F_{\rm equiv}^{(\hat a)}$ at the fixed probe length $\hat a=0.20$ for
the three non-reference swap probabilities. The mismatch is smallest for
$\ps=10^{-3}$ and increases systematically as $\ps$ is raised. This mirrors
the behavior of the direct clock differences in
Fig.~\ref{fig:signed_clock_T050}: weak swap produces only a small perturbation
of the reference dynamics, whereas stronger swap generates a progressively
larger departure from a common material clock.}

\mpc{The evolution with relaxation stage is also informative. 
Fig.~\ref{fig:deltaF_equiv}(a) shows that
$\delta F_{\rm equiv}^{(\hat a)}$ remains finite throughout the resolved
$\alpha$-relaxation window. Its variation with $F$ is weaker than that of the
raw clock difference $\Delta\log_{10}h$. Thus, the apparent decrease of the
clock mismatch in logarithmic-time units does not translate directly into an
equally strong reduction of its effect on the relaxation function.}

\mpc{Figure~\ref{fig:deltaF_equiv}(c) fixes the swap probability at $\ps=10^{-1}$ and compares
different probe lengths. The mismatch depends
systematically on $\hat a$, showing that the consequences of clock
non-uniqueness are themselves length-scale dependent. The failure of a common
clock therefore cannot be regarded merely as a numerical ambiguity in the
definition of material time: different probe-dependent clocks lead to
different predictions for the relaxation.}

\mpc{The inherent-structure results provide the same tests after suppressing fast
vibrational motion. Figures~\ref{fig:deltaF_equiv}(b) and (d) show that the
mismatch changes as structural relaxation proceeds,  while remaining finite and
probe-dependent over the accessible $\alpha$-relaxation window. Projection onto
inherent structures therefore does not restore an exact single-clock
description.}

\mpc{The clock mismatch identified in Fig.~\ref{fig:signed_clock_T050} thus has a
finite and measurable consequence for the relaxation functions themselves.
Its systematic dependence on both $\ps$ and $\hat a$ shows that, over the
resolved $\alpha$ regime, the effect of swap cannot be reduced to a single
reparameterization of time.}

\subsection{Distribution-level test and origin of the probe-length dependence}
\label{subsec:distribution_space}
\mpc{
The clock analysis above tests time reparameterization through a family of
collective overlaps measured at different probe lengths. A complementary and
stronger way to formulate the same question is to consider the underlying
distribution from which these overlaps are obtained.}

\mpc{
The idea is most transparent for the self overlap. The self overlap at probe
length $\hat a$ is the cumulative of the self part of the van Hove function
over displacements smaller than $\hat a$. Consequently, if two dynamics were
related by a reparameterization of time, then comparison at corresponding
relaxation stages should collapse not only the self overlap, but the full self
van Hove function. As shown in Appendix~\ref{app:beyond_mean},
Fig.~\ref{fig:self}, this does not occur: at equal self-relaxation stage the
shape of the displacement distribution still depends on the swap probability.
Thus, in the self sector, swap changes more than the time at which a given
amount of relaxation is reached.}

\mpc{
An analogous construction can be made for the collective overlap. For each
initial particle position $\mathbf r_i(0)$, we define the distance to the
nearest particle at time $t$,
\begin{equation}
d_i(t)=\min_j |\mathbf r_j(t)-\mathbf r_i(0)|,
\label{eq:dmin}
\end{equation}
and denote the corresponding probability
distribution by $P_{\min}(d,t)$. Unlike the self van Hove function,
$P_{\min}$ is identity blind: the particle found closest to an initial
position at time $t$ need not be the particle that occupied that position at
$t=0$.}

\mpc{
For the probe lengths considered here, $\hat a<l_0$, multiple occupation of
the probe volume is negligible. In this regime, the collective overlap is
determined by the cumulative probability that the nearest particle lies within
the probe radius,
\begin{equation}
\Qtot(\hat a,t)
=
\frac{
\displaystyle \int_0^{\hat a} P_{\min}(d,t)\,{\rm d}d-q_r(\hat a)
}{
1-q_r(\hat a)
}.
\label{eq:pmin_overlap}
\end{equation}
The collective overlaps at different $\hat a$ therefore correspond to
cumulative integrals of the same underlying distribution $P_{\min}(d,t)$,
evaluated at different upper limits.}

\mpc{
This immediately yields a distribution-level test of time
reparameterization. If swap only changed the material time, then a single
clock $h_{\ps}(t)$ should align not only the overlaps, but the full
identity-blind distribution,
\begin{equation}
P_{\min}^{(\ps)}(d,t)
\simeq
P_{\min}^{({\rm ref})}(d,h_{\ps}(t)).
\label{eq:pmin_reparam}
\end{equation}
Collapse of a finite set of overlaps is therefore a weaker condition than
collapse of the full distribution from which they are constructed.}

\begin{figure}[!t]
  \includegraphics[width=1.0\textwidth]{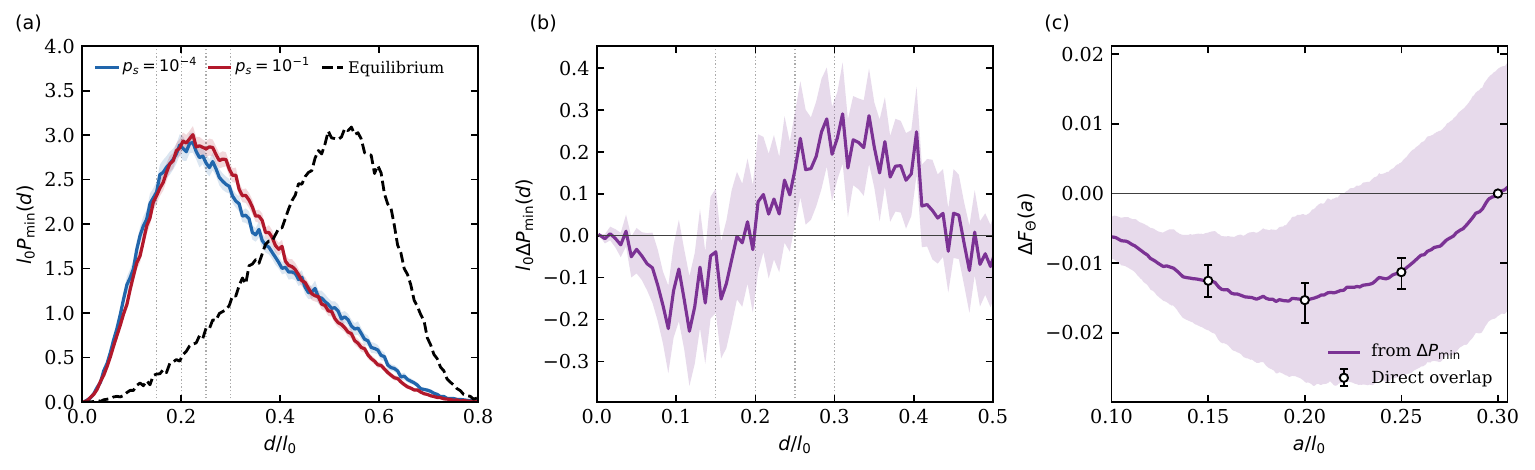}
  \caption{Distribution-level origin of the probe-length dependence at
  $T=0.50$. Reference dynamics ($p_s=10^{-4}$) and strong-swap dynamics
  ($p_s=0.1$) are compared at the matched collective relaxation stage
  $F_\Theta(0.30l_0)=0.5$.
  (a) Nearest-distance distributions $P_{\min}(d,t)$ for the two dynamics,
  together with the independently measured equilibrium distribution reached
  when the initial and final configurations are decorrelated. Vertical dotted
  lines indicate the probe radii $\hat a=0.15$, $0.20$, $0.25$, and $0.30$.
  The reference and swap distributions remain distinct, and both are still far
  from their common equilibrium limit.
  (b) Difference
  $\Delta P_{\min}(d)=P_{\min}^{\rm swap}(d)-P_{\min}^{\rm ref}(d)$.
  Its change of sign shows that equality of one cumulative overlap can result
  from cancellation between distinct portions of the underlying distribution.
  (c) Corresponding cumulative overlap difference
  $\Delta F_\Theta(\hat a)$ obtained from Eq.~\eqref{eq:deltaF_from_pmin}.
  Because the two dynamics are matched at $\hat a_0=0.30$, the cumulative
  difference vanishes there, whereas it remains finite at smaller probe
  lengths. Shaded bands and point error bars are pointwise $95\%$ confidence intervals
from $1000$ independent-trajectory bootstrap resamples, performed separately
for the two dynamics.}
  \label{fig:pmin_decomposition}
\end{figure}

\mpc{
Figure~\ref{fig:pmin_decomposition} shows this comparison for the reference
dynamics ($p_s=10^{-4}$) and strong swap ($p_s=0.1$) at the matched collective
relaxation stage $F_\Theta(\hat a_0)=0.5$, with $\hat a_0=0.30$.
Panel~(a) shows the nearest-distance distributions themselves, together with
the independently measured equilibrium distribution obtained from statistically
independent configurations. 
The reference and swap distributions do not collapse.
Since both distributions remain far from their common equilibrium limit reached at times proportional but much larger than $\tau_\alpha$~\cite{CommentBerthier},
their increasing similarity at late stages cannot be attributed simply to
having approached the trivial long-time limit.}

\mpc{The same figure also clarifies the origin of the probe-length dependence.
Defining
\begin{equation}
\Delta P_{\min}(d)
=
P_{\min}^{(\ps)}(d)-P_{\min}^{({\rm ref})}(d),
\label{eq:delta_pmin}
\end{equation}
the corresponding difference in collective overlap at probe $\hat a$ is
\begin{equation}
\Delta F_\Theta(\hat a)
=
\frac{
\displaystyle \int_0^{\hat a} \Delta P_{\min}(d)\,{\rm d}d
}{
1-q_r(\hat a)
}.
\label{eq:deltaF_from_pmin}
\end{equation}
Panel~(b) shows that $\Delta P_{\min}(d)$ changes sign within the
probe-relevant range. Consequently, different probe radii integrate different
portions of the same non-collapsing distribution. Panel~(c) shows the resulting
cumulative mismatch $\Delta F_\Theta(\hat a)$: because the two dynamics are
matched at $\hat a_0=0.30$, positive and negative contributions cancel at that
radius, whereas the cancellation is incomplete at smaller $\hat a$.}

\mpc{
This directly explains the existence of a probe-dependent collective clock.
For the stages considered here, $\Delta F_\Theta(\hat a)$ is negative at the
smaller probe lengths, consistent with the positive
$\Delta\log_{10}h_{\ps}^{(\hat a)}$ reported above, since the collective
overlaps decrease with time. The cumulative mismatch also reproduces the broad
ordering of the probe dependence, with the smallest deviation at the probe
closest to the matching length. The more detailed ordering of the smallest
probes is not determined by the cumulative difference alone, because
converting a vertical mismatch in $F_\Theta$ into a horizontal mismatch in the
clock also involves the probe-dependent local slope of the relaxation curve.}

\mpc{
We therefore conclude that the approximate collapse of the collective overlaps
does not reflect an exact collapse of the underlying collective displacement
distribution. Rather, different probe lengths sample different cumulative
projections of a distribution that remains swap dependent.}

\section{Discussion}

We have tested the time-reparameterization interpretation of swap Monte Carlo
dynamics by making explicit the clock implicit in collective-overlap
parametric plots. We first reproduced the empirical observation of
Ref.~\citenum{ghimenti2026}: mean collective density overlaps at different probe
lengths approximately collapse when time is eliminated. We then showed that
this collapse does not define a unique material time.

A strict time reparameterization requires one clock $h_{\ps}(t)$ for each
dynamics, independent of the observable used to extract it. This single-clock
condition is violated even within the mean collective sector. Clocks extracted
from different collective probe lengths are systematically different: their
signed differences are ordered by probe length, depend on the relaxation stage,
and are statistically resolved over a substantial part of the slow window.
\mpc{A global compromise clock captures the dominant acceleration, but still
leaves structured probe-length-dependent residuals.}
The robustness of this conclusion is discussed in
Appendix~\ref{app:robustness}.

This result clarifies what is established by Ghimenti-style parametric plots.
Those plots show that different dynamics follow similar projected curves in the
space of mean collective overlaps, which is a useful and nontrivial observation.
However, pairwise parametric collapse is weaker than single-clock
reparameterization: a pairwise plot can look accurate even when the material
time inferred from different probe lengths differs systematically. The data
therefore support an approximate projected mean-overlap reparameterization, not
an exact observable-independent clock relating swap and \mpc{the reference
dynamics}.

\mpc{
The distribution-level analysis makes the same distinction more directly.
The collective overlaps at different probe lengths are cumulative projections
of the same nearest-distance distribution $P_{\min}(d,t)$. At matched
collective relaxation stages, the reference and strong-swap distributions
remain distinct, while positive and negative differences partially cancel when
integrated over the probe volume. The good collapse of selected collective
overlaps can therefore coexist with a measurable difference in the underlying
distribution. This provides a microscopic origin for the probe-length
dependence of the extracted clock.
}

\mpc{
We observe that the clock
mismatch decreases on cooling: at a fixed relaxation stage
$\Qtot(\hat a_0,t)=0.65$ and $\ps=10^{-1}$, the largest probe-length clock
difference decreases from $\Delta\log_{10}h\simeq0.1$ at $T=0.50$ to
$\simeq0.05$ at $T=0.45$, within the high-overlap window accessible at the
lower temperature, as discussed in Appendix~\ref{app:robustness}.
This observation is consistent with time-reparameterization invariance in aging systems, which is expected to become exact only asymptotically, as the dynamics becomes infinitely
slow~\cite{Kennett2001,Chamon2002,Kurchan2023}. 
We therefore do not claim that reparameterization invariance is violated in the asymptotic limit, which our simulations cannot reach. Rather, we find that a unique
observable-independent clock has not emerged over the numerically accessible
$\alpha$-relaxation regime.
}

\mpc{
Our results concern one representative continuously polydisperse glass former
and therefore do not establish how the magnitude of the clock mismatch varies
across models. They do, however, provide a counterexample to the stronger claim
that the success of collective parametric collapse generally implies an exact
single material clock for swap dynamics. Establishing whether the mismatch
vanishes asymptotically, or how rapidly it decreases in other glass formers,
requires further tests at larger time-scale separation.
}

This distinction matters for the proposed reconciliation between
landscape-based and kinetic descriptions. The time-reparameterization scenario
suggests that kinetic rules determine the clock, while the static landscape
determines reparameterization-invariant slow collective relations.
\mpc{Our results show that this separation is only approximate: the mean
collective-overlap relations are nearly invariant, whereas the material clock
remains systematically probe-dependent. Thus the swap-efficiency challenge to
a purely thermodynamic control of the physical relaxation time is not
eliminated by the observed collective mean-overlap reparameterization.}

\appendix

\section{Beyond mean collective overlaps}
\label{app:beyond_mean}

The main text focuses on the mean collective-overlap sector, where the
time-reparameterization picture appears most successful. Here we briefly
record two complementary checks showing why this collective clock should not
be promoted to a full dynamical equivalence. A change of time can shift when a
curve is reached, but it cannot change fluctuation amplitudes or the shape of
a displacement distribution when configurations are compared at the same
relaxation stage.

\begin{figure}[!t]
  \includegraphics[width=1\textwidth]{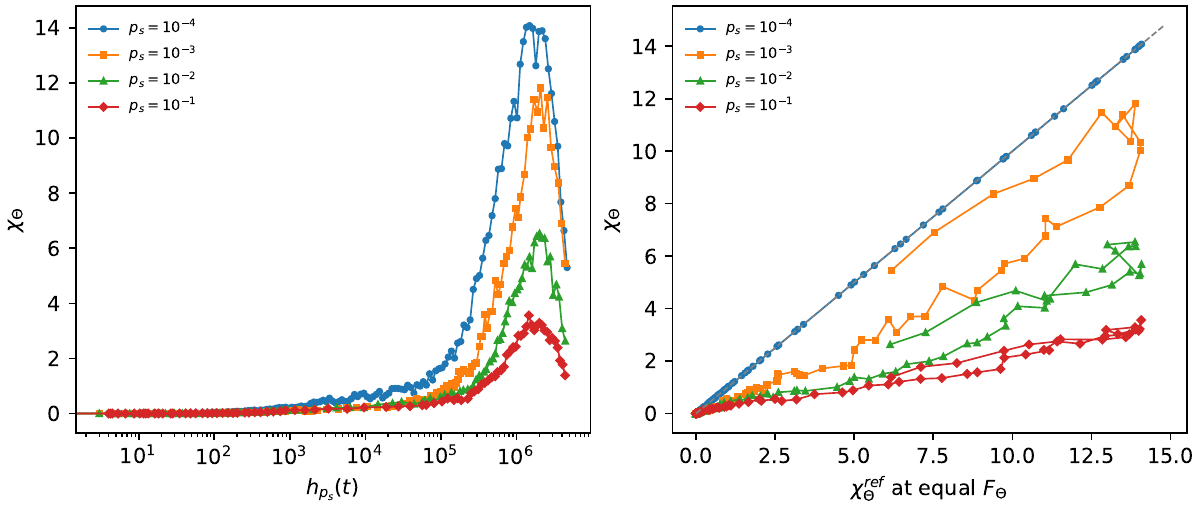}
  \caption{The collective mean clock does not collapse the fluctuations.
  For each swap probability $\ps$ we extract the clock that maps the mean
  overlap $\Qtot(\hat a=0.3,t\mid\ps)$ onto the reference dynamics
  ($\ps=10^{-4}$). (a) The collective-overlap susceptibility $\chi_\Theta$
  plotted against this mean-derived clock: the peaks occur at comparable
  relaxation stages but their amplitudes remain ordered by $\ps$ and are
  suppressed as swap increases. (b) At equal mean overlap,
  $\chi_\Theta(\ps)$ versus the reference $\chi_\Theta^{\rm ref}$: every swap
  dynamics lies systematically below the diagonal. Data at $T=0.50$,
  $\hat a=0.3$.}
  \label{fig:chi4}
\end{figure}

Figure~\ref{fig:chi4} shows that the clock extracted from the mean collective
overlap does not collapse the corresponding fluctuations. The susceptibility
peaks remain strongly dependent on $\ps$, and their amplitudes decrease as the
swap rate increases. This is expected: a reparameterization of the time axis
can move the peak position, but it cannot restore the peak amplitude at a
fixed stage of relaxation.

\begin{figure}[!t]
  \includegraphics[width=1\textwidth]{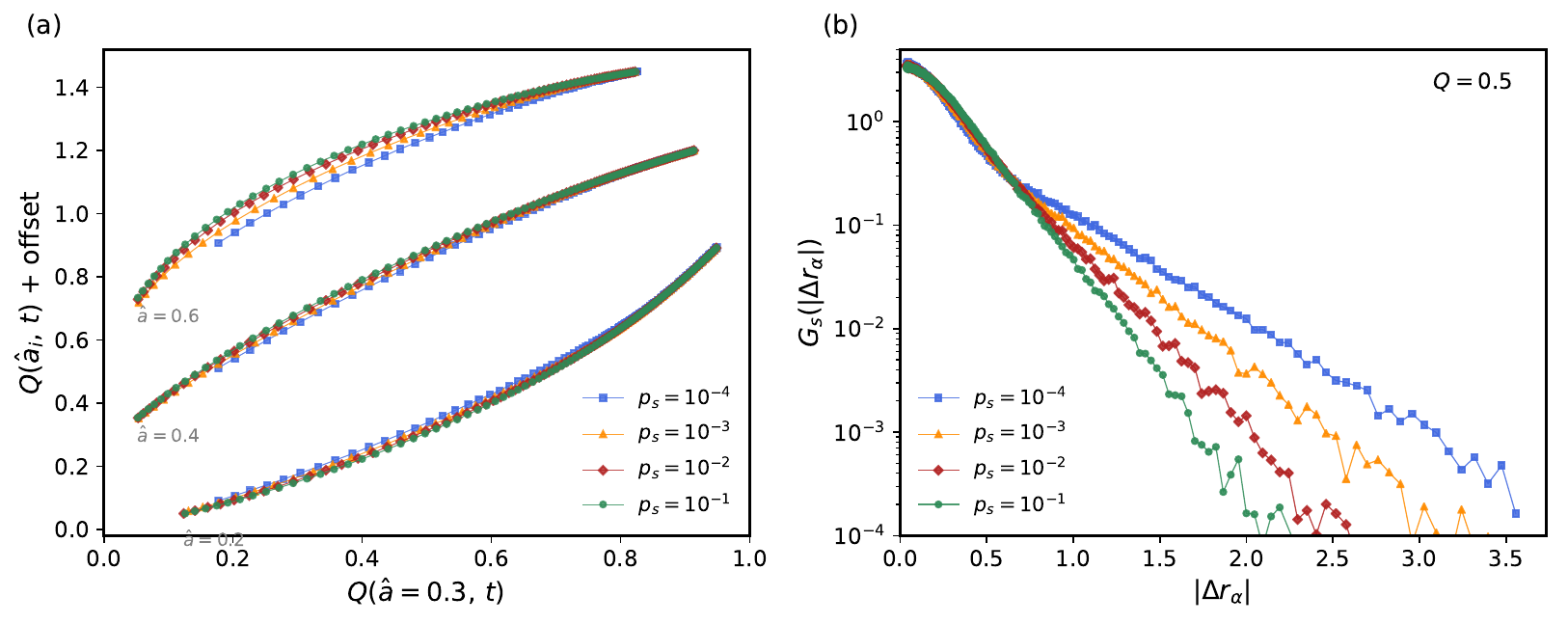}
  \caption{Particle-resolved observables are not
  reparameterization-invariant. (a) Parametric plot of the self-overlap
  $\Qself(\hat a_i,t)$ against the reference $\Qself(\hat a=0.3,t)$ at three
  probe lengths, for several swap probabilities $\ps$. Unlike the collective
  mean overlap, the self-overlap curves do not collapse. 
  (b) \mpc{
  Distribution $G_s(\Delta r;t)$ of the absolute Cartesian displacement
  component, compared at the fixed relaxation stage
  $Q(\hat a=0.30,t)=0.5$, for several $p_s$. Displacements are reported in units of $\sigma_{\min}=1$.}
   Increasing the swap rate
  progressively suppresses the non-Gaussian tail of large displacements. Data
  at $T=0.50$.}
  \label{fig:self}
\end{figure}

Figure~\ref{fig:self} shows the analogous limitation at the particle-resolved
level. The self-overlap does not display the same parametric collapse as the
collective overlap, and the van Hove function changes shape at fixed
relaxation stage. These observations are consistent with the known suppression
of dynamic heterogeneity by swap. They are not the central test of the present
work, but they delimit the scope of the collective reparameterization: the
approximate clock is a property of the projected mean collective relaxation,
not of the full dynamics.

\section{A global compromise clock leaves structured residuals}
\label{app:globalclock}

The previous test extracts a clock independently at each probe length. To
construct a single compromise clock, we define
\begin{equation}
\log_{10} h_{\ps}^{\rm glob}(t)
=
\frac{1}{N_{\hat a}}
\sum_{\hat a}
\log_{10} h_{\ps}^{(\hat a)}(t).
\end{equation}
At each time $t$, this choice minimizes the squared dispersion of the
probe-dependent clocks in $\log_{10}h$. Since the individual clocks are
monotonic functions of $t$, their average also defines a monotonic global
clock. We then compare the swap overlaps
$\Qtot^{(\ps)}(\hat a,t)$ with the reference overlaps evaluated at this clock,
$\Qtot^{({\rm ref})}(\hat a,h_{\ps}^{\rm glob}(t))$.

\begin{figure}[!t]
  \includegraphics[width=1.0\textwidth]{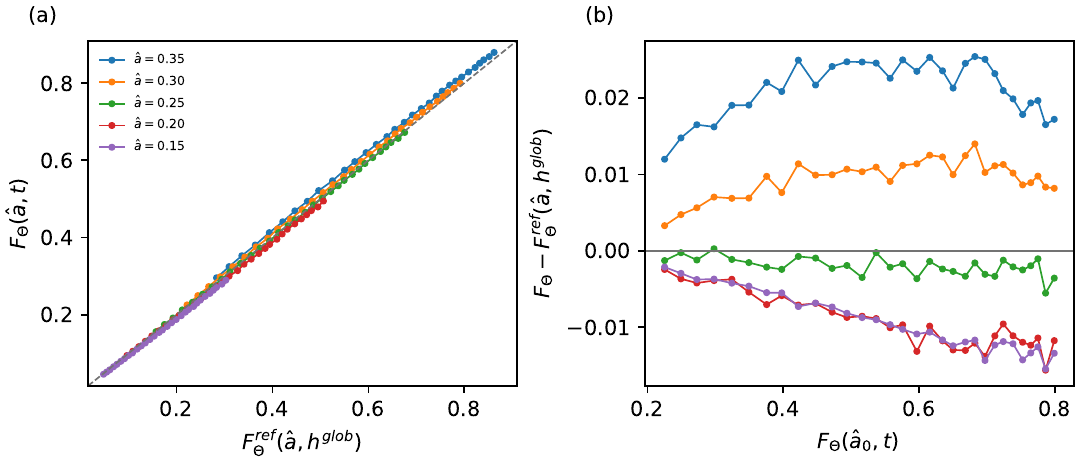}
  \caption{Global-clock test at $T=0.50$ for $\ps=10^{-1}$.
  (a) Swap overlaps $\Qtot^{(\ps)}(\hat a,t)$ plotted against the corresponding
  reference overlaps evaluated at $h_{\ps}^{\rm glob}(t)$. The diagonal
  indicates perfect single-clock collapse. (b) Residuals
  $\Qtot^{(\ps)}(\hat a,t)-
  \Qtot^{({\rm ref})}(\hat a,h_{\ps}^{\rm glob}(t))$, shown as a function of
  the relaxation stage $\Qtot^{(\ps)}(\hat a_0,t)$, with $\hat a_0=0.30$.
  The global clock captures the dominant acceleration, but leaves systematic
  probe-dependent residuals.}
  \label{fig:global_clock_T050}
\end{figure}

Figure~\ref{fig:global_clock_T050} shows that the global clock captures the
dominant relaxation trend, but the residuals remain ordered by probe length
and change sign across the set of probes. The failure of clock uniqueness is
therefore not an artifact of extracting the clocks independently.

The vertical residuals are small, typically of order $10^{-2}$, but are large
compared with the statistical uncertainty of the measured overlaps. 
\mpc{As a
descriptive measure, we define
\begin{equation}
R_{\hat a}
=
\left\langle
\left[
\frac{
\Qtot^{(\ps)}(\hat a,t)
-
\Qtot^{({\rm ref})}(\hat a,h_{\ps}^{\rm glob}(t))
}{
{\rm SE}[\Qtot^{(\ps)}(\hat a,t)]
}
\right]^2
\right\rangle_t^{1/2},
\end{equation}
where the average is over $0.2\leq\Qtot^{(\ps)}(\hat a_0,t)\leq0.8$ and
${\rm SE}$ indicates the standard error.
For strong swap, the largest values are $R_{\hat a}\simeq8.3$ for the thermal
trajectories and $\simeq6.3$ for the inherent structures. These values do not
include propagation of the uncertainty in the inferred global clock.}

Thus both the independent-clock and global-clock analyses support the same
conclusion: the collective overlaps admit an accurate approximate
reparameterization, but not an exact single material clock.

\section{Robustness of the collective-clock analysis}
\label{app:robustness}

The main text presents the single-clock analysis at $T=0.50$, where the
collective overlaps relax over the full slow window used for the clock
construction. Here we report two complementary checks. 
First, we summarize the size and statistical resolution of the clock mismatch at $T=0.50$. Second, we repeat the analysis at the lower temperature $T=0.45$. 
Since this temperature lies below the mode-coupling crossover of the model, $T_c\simeq0.50$, the reference and weak-swap dynamics are very slow on the time scales accessible to
our simulations. 
As a result, the overlaps do not fully decorrelate, and the clock analysis at
$T=0.45$ is restricted to the accessible high-overlap part of the relaxation,
roughly $F_\Theta(\hat a_0,t)\simeq0.6$--$0.8$, \mpc{using only the overlap range
jointly covered by the reference and swap curves, without extrapolation}.
The $T=0.45$
results should therefore be read as a consistency check in the early part of
the slow decay, not as a full alpha-relaxation analysis.

\mpc{
To summarize the size and statistical significance of the clock mismatch at
$T=0.50$, we define
\begin{equation}
M(t)=
\max_{\hat a}
\left|
\Delta\log_{10} h_{\ps}^{(\hat a)}(t)
\right|.
\end{equation}
The uncertainty $\sigma_M(t)$ is taken as the standard deviation of $M(t)$
over the same trajectory-level bootstrap realizations used to construct the
confidence bands in Fig.~\ref{fig:signed_clock_T050}. The ratio
$M(t)/\sigma_M(t)$ therefore provides a compact measure of the statistical
resolution of the clock non-uniqueness.
}

\begin{figure}[!t]
  \includegraphics[width=1.0\textwidth]{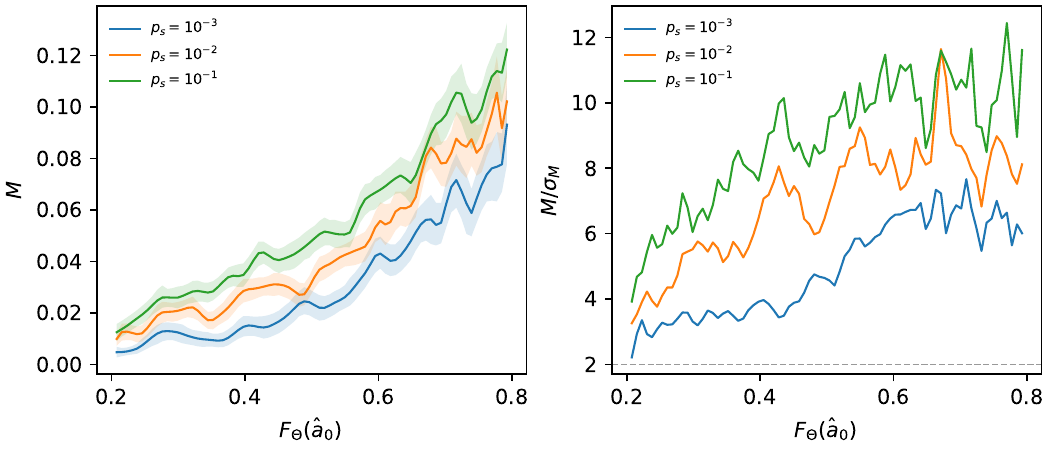}
  \caption{Summary of the clock mismatch at $T=0.50$. The maximum mismatch
  $M(t)=\max_{\hat a}|\Delta_{\hat a}(t)|$ and the corresponding
  signal-to-noise ratio $M(t)/\sigma_M(t)$ are shown as functions of the
  relaxation stage. This representation compresses the probe-length dependence
  shown in Fig.~\ref{fig:signed_clock_T050} into a single amplitude and
  confirms that the clock non-uniqueness is statistically resolved over a
  substantial part of the slow relaxation window. 
  \mpc{The dashed horizontal line at $M/\sigma_M=2$ is included as a visual reference.}}
  \label{fig:mismatch_summary_T050}
\end{figure}

\begin{figure}[!t]
  \includegraphics[width=1.0\textwidth]{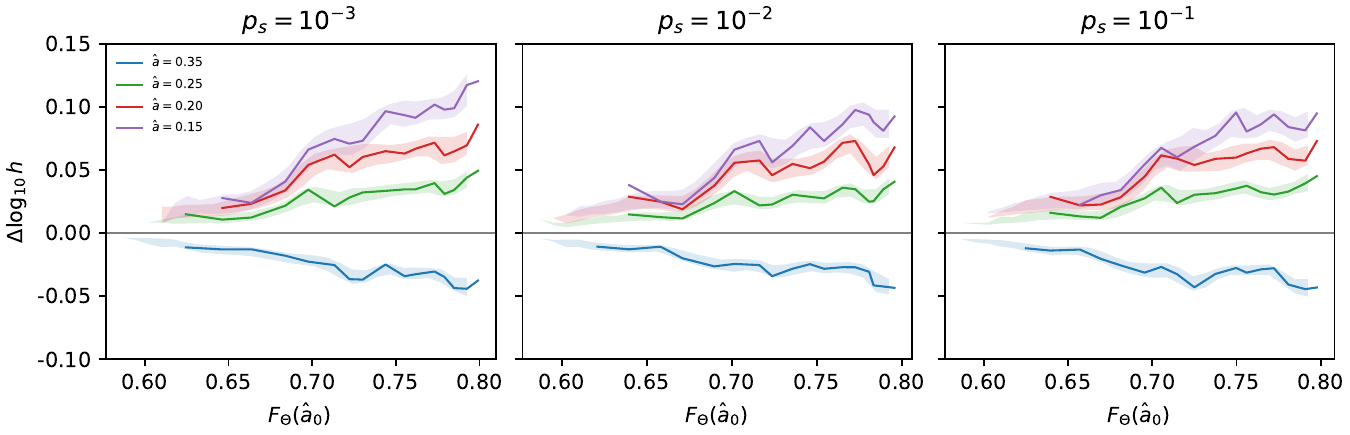}
  \caption{Signed clock differences at $T=0.45$. As in
  Fig.~\ref{fig:signed_clock_T050}, clocks extracted from different probe
  lengths are compared to the clock extracted from $\hat a_0=0.30$. Because
  $T=0.45$ lies below $T_c\simeq0.50$, the dynamics is not fully relaxed over
  the accessible simulation window; the analysis is therefore restricted to the
  high-overlap part of the slow decay. Within this accessible range, the clock
  differences remain ordered by probe length.}
  \label{fig:signed_clock_T045}
\end{figure}

\begin{figure}[!t]
  \includegraphics[width=1.0\textwidth]{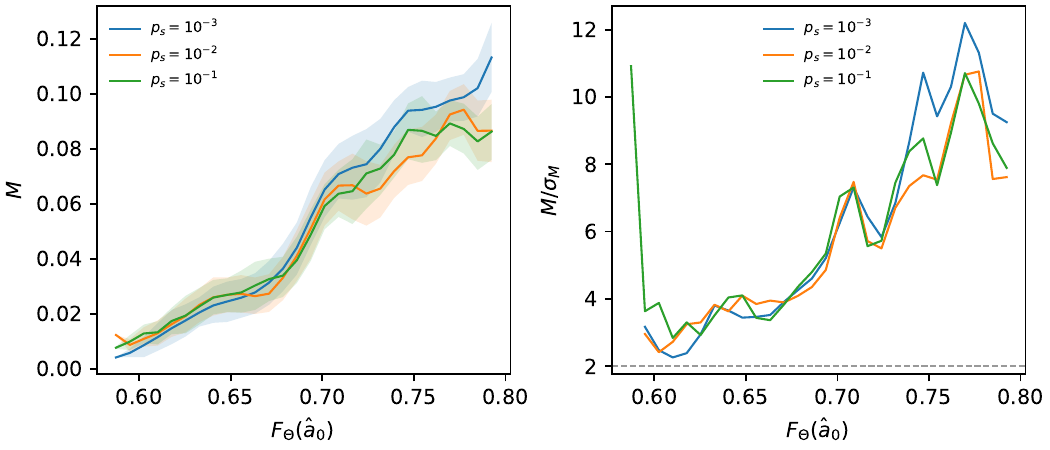}
  \caption{Summary of the clock mismatch at $T=0.45$, analogous to
  Fig.~\ref{fig:mismatch_summary_T050}. Because relaxation is incomplete at
  this lower temperature, the plotted range is limited to the accessible
  high-overlap window. The mismatch remains statistically resolved within this
  range.}
  \label{fig:mismatch_summary_T045}
\end{figure}

\begin{figure}[!t]
  \includegraphics[width=1.0\textwidth]{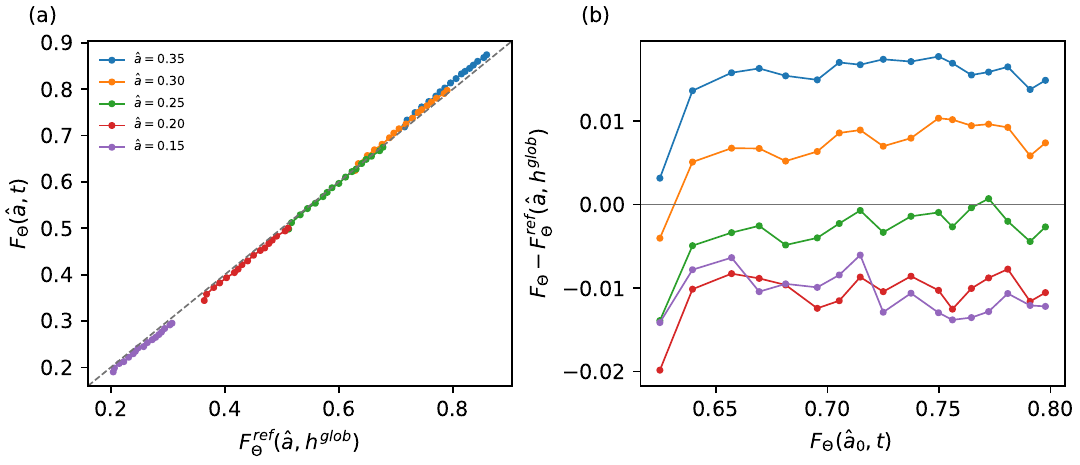}
  \caption{Global-clock test at $T=0.45$. The analysis is restricted
  to the accessible high-overlap part of the relaxation. As at $T=0.50$, the
  best global clock captures the dominant trend but leaves structured
  probe-length-dependent residuals.}
  \label{fig:global_clock_T045}
\end{figure}

Figure~\ref{fig:mismatch_summary_T050} summarizes the main-text result at
$T=0.50$: the probe-length dependence of the clock is not only systematic but
also statistically resolved over a substantial part of the slow window. 
The lower-temperature data provide a useful, but more limited, check, as the dynamics is so slow that the simulations do not reach the full decay of $\Qtot(\hat a_0,t)$. 
Consequently, Figs.~\ref{fig:signed_clock_T045}--\ref{fig:global_clock_T045} cover only the
accessible high-overlap range. Within that range, the same qualitative
signatures are observed: the signed clock differences are ordered by probe
length (Fig.~\ref{fig:signed_clock_T045}), the maximum mismatch remains
statistically resolved (Fig.~\ref{fig:mismatch_summary_T045}), and a global compromise clock leaves structured residuals (Fig.~\ref{fig:global_clock_T045}). Thus the
clock non-uniqueness observed at
$T=0.50$ is not a peculiarity of that temperature. The $T=0.45$ data support
the same interpretation, while the incomplete relaxation prevents an equally
stringent test over the full alpha-relaxation window. At matched relaxation
stage, the mismatch amplitude is also smaller at $T=0.45$ than at $T=0.50$,
consistent with the collective reparameterization sharpening as the dynamics
slows.

\begin{acknowledgments}
This research was supported by the Singapore Ministry of Education through grants
MOE-T2EP50221-0016 and RG152/23.
\end{acknowledgments}

\bibliographystyle{aipnum4-1}
\bibliography{references}

\end{document}